\documentclass[usenatbib]{mn2e}
\usepackage{times}
\usepackage{graphicx}
\usepackage[ansinew]{inputenc}
\usepackage[T1]{fontenc}
\usepackage{lmodern}
\usepackage{amssymb,amsmath}
\usepackage{rotating}
\usepackage{multirow}
\usepackage{pdflscape}
\numberwithin{equation}{section}

\title[The Lutz-Kelker Paradox ]{The Lutz-Kelker Paradox}
\author[Charles Francis]{Charles Francis$^{1}$\thanks{E-mail: C.E.H.Francis.75@cantab.net} \\
$^{1}$25 Elphinstone Rd., Hastings, TN34 2EG, UK.}

\begin{document}

\pagerange{\pageref{firstpage}--\pageref{lastpage}} \pubyear{2014}

\maketitle

\label{firstpage}

\begin{abstract}
The Lutz-Kelker correction is intended to give an unbiased estimate for stellar parallaxes and magnitudes, but it is shown explicitly that it does not. This paradox results from the application of an argument about sample statistics to the treatment of individual stars, and involves the erroneous use of a frequency distribution in the manner of a probability density function considered as a Bayesian prior. It is shown that the Bayesian probability distribution for true parallax given the observed parallax of a selected star is independent of the distribution of other stars. Consequently the Lutz-Kelker correction should not be used for individual stars. This result has important implications for the RR Lyrae scale and for the interpretation of results from Gaia and Hipparcos. The Lutz-Kelker correction is a poor treatment of the Trumpler-Weaver bias which affects parallax limited samples. A true correction is calculated using numerical integration and confirmed by a Monte Carlo method.
\end{abstract}


\section{Background}\label{sec:1}
The issue of statistical bias in astronomical analysis was addressed as long ago as 1913, by Eddington, who considered the bias to the number of observed stars within a given range of magnitudes caused by observational errors in magnitude. Indeed, whenever a stellar sample is selected either by magnitude or by a bound on parallax there will be bias to the statistics of that population. Consequently, a correction will need to be made dependent upon the sample selection. Distinct from bias to statistics, Lutz and Kelker (1973, hereinafter ``LK1'') sought to address bias resulting from observations of individual stars.

The question of bias to distances and magnitudes derived from parallax measurements has gained in importance since Hipparcos and with accurate determinations of parallaxes of near RR Lyrae by the Hubble Space Telescope (HST, Benedict et al. 2011) and it will be vital to the correct interpretation of data from Gaia. Selection bias can affect observations of individual stars as well as samples. If, for example, the brightest star (or stars) of a given class has been chosen, then it may not be representative of the class. Groenewegen (2008) and Francis and Anderson (2014) identified a selection bias in calibrations of red clump magnitudes based on a sample for which $I$ band measurements are available. Similarly one should consider whether other magnitude calibrations from single stars or from small samples should be subject to bias. 

A correct treatment may depend upon the particular properties of the population under consideration. RR Lyrae itself is brighter than other RR Lyrae variables by more than a magnitude (Fernley et al. 1998). We can conclude that it is sufficiently close that it would be the brightest RR Lyrae irrespective of whether it is atypical in any way. Therefore no bias should be assumed. However, the selection of further RR Lyrae variables is inevitably biassed towards bright stars so as to select near stars and reduce parallax errors. Consequently a Malmquist (1920) correction should be applied. Because of other selection criteria, and because of uncertainties in the Malmquist correction itself, it is probably not correct to apply the full Malmquist correction of $ \sim 0.014 $ mag assuming an intrinsic dispersion in magnitude $ \sigma \approx 0.1. $  

Because of the large size of the Malmquist bias for many stellar classes, and the difficulty in estimating it accurately, it is preferable when possible to work with samples limited by parallax or by parallax error. This introduces a different range of problems, which are the main topic of this paper. As reported in a review by Smith (2003), the Lutz-Kelker (L-K) bias has been a source of confusion since it was originally described. There has been some doubt in the minds of many astronomers as to whether the bias is real, whether it is correctly calculated in published treatments, and when it should be applied. Whereas Sandage and Saha (2002) state \textit{``The Lutz-Kelker paper was so clear that it soon became the principal reference to the problem''}, Smith (2003) finds that \textit{``the seeming clarity of LK1 masks a profound ambiguity surrounding the nature of the Lutz-Kelker bias''}. I will show here that the implications are more serious than Smith realised.

Apparently a lack of clarity continues, because Feast (2002) states that \textit{``when only one object of a class is measured, as in the important Hubble Space Telescope (HST) observations of the parallaxes of RR Lyrae and $ \delta $ Cephei (Benedict et al. 2002a,b)''} that \textit{``the absolute magnitudes derived from the parallaxes are therefore not subject to Lutz-Kelker bias''}, but Smith (2003) describes the approach taken by LK1 as \textit{``implicitly Bayesian''} and Benedict et al. (2011) applied a significant L-K correction, saying that although LK1's original argument was frequentist, it can be put into a Bayesian form and applied to parallaxes of individual stars. 

This letter seeks to make clear that, regardless of whether the argument is Bayesian or frequentist, Feast is correct. The L-K correction should not be applied to parallaxes of individual stars. The issue is not one of the philosophical difference between Bayesian and frequentist analysis; in either case one requires that the mean of many measurements should not be biased. However, there is a subtle error in LK1's argument; it is neither frequentist nor Bayesian, but is based on an incorrect use of Bayes' theorem, perhaps because it was formulated before Bayesian statistics were widely recognised and used. 

Because of the importance of this result, I have shown it in a number of ways, by demonstrating a contradiction from the probability distribution proposed by LK1, by mathematical argument to show the correct probability distribution, and by comparison of the results of numerical integration with Monte Carlo models simulating stellar distribution. It is seen that the correction to distance and magnitudes due to errors in parallax for individual stars is much smaller than, and of the opposite sign, to the correction calculated by Lutz and Kelker, and that the Trumpler-Weaver bias for parallax limited samples is smaller than the supposed L-K bias, and cannot be identified with it.

\section{Lutz and Kelker's argument}\label{sec:2}
LK1 began by describing the Trumpler-Weaver (1953) bias. This is a selection bias affecting the mean parallax distance of a population within a sphere of given radius. The number of stars with true distances greater than this radius which appear in the sample due to parallax error will exceed the number with true distances inside the sphere whose parallax errors remove them from the sample, because the volume of the error shell outside the sphere is greater than the volume of the error shell inside the sphere. The consequence is that the true mean distance of the sample is greater than the mean of observed parallax distances.

They then considered stars at the surface of this sphere, observing that more stars observed to have parallax $\pi_0$ are actually outside the sphere than inside it, so that the average true parallax for stars with observed parallax $\pi_0$ is smaller than $\pi_0$. They assumed a Gaussian probability distribution for the observed parallax, $\pi_0$, about the true parallax, $\pi$, 
\begin{equation}
f(\pi_0|\pi) = N(\pi_0, \pi, \sigma),
\end{equation}
and asserted a probability distribution for the true parallax given the observed parallax, 
\begin{equation}
g(\pi|\pi_0) \propto (1/\pi)^4 N(\pi, \pi_0, \sigma),
\end{equation}
which takes into account that the frequency density of stars is proportional to $ r^2 $. Writing $ Z = \pi/\pi_0 $ they found
\begin{equation}
g(\pi|\pi_0) \propto G(Z) \equiv Z^{-4} N(Z, 1, \sigma/\pi_0).
\end{equation}
They evaluated the bias in magnitude as the expectation of
\begin{equation}
\Delta M = M_{\mathrm{true}} - M_{\mathrm{observed}} = 5\log(\pi/\pi_0) = 5\log(Z),
\end{equation}
giving 
\begin{equation}
\langle \Delta M(\epsilon) \rangle = \frac{5 \int_{\epsilon}^{\infty} \log Z G(Z) dZ}{ \int_{\epsilon}^{\infty} G(Z) dZ},
\end{equation}
where a small value $ \epsilon $ was introduced because the integrals are not defined in the limit $ \epsilon\rightarrow 0 $. They observed that the resulting expression for $ \langle \Delta M(\epsilon) \rangle $ in (2.5) is also divergent in the limit $ \epsilon\rightarrow 0 $. However, they argued (a) that this is not important because a near constant result is found for a range of small values of $ \epsilon $, and (b) that the limit $ \epsilon\rightarrow 0 $ is not physically relevant because stars with a large observational error such that  $ \pi/\pi_0 $ is small are easily recognised and excluded. Since $ \langle \Delta M(\epsilon) \rangle $ is not explicitly dependent on parallax, they claimed that the bias is present in all measured parallaxes.

\section{The Paradox}\label{sec:3}
As presented in the LK1, the argument described in section 2 appears both reasonable and clear, so much so that it is difficult to think it could be in error. However, this apparent clarity disguises a number of mathematical and logical problems, and it is perhaps better considered as among the most subtle paradoxes in the history of physical science (of which there are many). The paradox is this. If we take many measurements of a star with true parallax $\pi'$, finding a large number of values of $\pi_0$ then the expectation of $\pi$ using distribution in (2.2) should be an unbiased estimator for $\pi'$. But in fact $ E(\pi) \neq \pi' $ (appendix A). Since it is easily seen that the mean of the observed parallaxes is an unbiased estimator for $\pi'$, we have to conclude that LK1's proposed distribution (2.2) is not consistent with their assumption (2.1). It is necessary to look at their argument more carefully.

LK1 argued that the lower bound, $ \epsilon $, could be introduced into (2.5) because small values of $Z$ are not physically important. But this undermines their claim that \textit{``The bias is not caused by the use of a lower parallax limit. Instead it exists at all values of parallax''}. In practice, when true parallaxes are sufficiently small, stars with zero and even negative parallaxes are recorded. In fact, if we assume a frequency density proportional to  $ r^2 $, the distribution will be dominated by stars with small parallaxes. Consequently both the factor $ Z^{-4} $ in $ G(Z) $ and the logarithm in the numerator lead to pathological behaviour of the integrals in (2.5) as $ \epsilon\rightarrow 0 $, including pathological behaviour of the ratio.

Certainly stars with sufficiently small parallaxes will be excluded from analysis, but then LK1's claim that there is no lower bound on parallax is false. Indeed, a lower bound on parallax must be used because it is axiomatic in probability theory that the integral of a probability density function is unity (this is not merely a convention as stated in LK1 but is fundamental to the mathematical definition of probability). It is meaningless to use a non-normalisable function as a probability density. The lower bound $ \epsilon $ used by LK1 simply replaces a well defined bound in terms of the observed parallax, $\pi_0$, or the error, $\sigma/\pi_0 $, with a bound defined from an unknown quantity, the true parallax, $\pi'$. It is a general corollary to the Riemann rearrangement theorem (e.g., Bromwich 1992) that one cannot simply insert a bound and achieve an unambiguous result, because the result depends upon the way in which the limit is taken. A consequence seen in section 4 is that the L-K correction overestimates the effect of the Trumpler-Weaver bias.

The problem originates in LK1's attempt to apply the method used by West (1969) to the case of a single star using a probability distribution in the form of (2.2). (2.2) cannot be normalised and cannot be treated as a probability distribution. West was not concerned with single stars, but with the mean systematic error for a collection of stars, as described by Trumpler and Weaver. If one selects a star at random from a given population, then the probability distribution for the parallax is proportional to the frequency distribution for the parallaxes of the population. This is not the same as the probability distribution for the parallax of a single pre-selected star. LK1 conflated these distinct ideas.

One cannot give a simple frequentist meaning to the probability distribution of the true parallax of a particular star given the observed parallax because the true value of the parallax of a star is a fixed quantity for that star, independent of the distribution of other stars, and does not have a range of different values in different trials. The LK1 analysis would only apply if stars were selected at random, irrespective of distance or magnitude, or if we actually did select stars to have a particular value of trigonometric parallax. Neither of these is ever done.

A valid probability distribution, with both Bayesian and frequentist meanings, can be given for the errors, or differences $ x = \pi_0 - \pi $ between observed and true parallaxes. For convenience, I will assume that there is a single best estimate of the observed parallax. In practice this is so because when many independent measurements are made we reduce them to a single estimate, given by the mean and the standard error of the mean. In accordance with (2.1) we assume the error distribution for the best estimate is Gaussian with mean zero, 
\begin{equation}
f(x) = N(x, 0, \sigma).
\end{equation}
The error, $x$, is independent of the estimate. We may therefore immediately conclude that the Bayesian probability distribution for the true parallax given the observed parallax is
\begin{equation}
f(\pi|\pi_0) = N(\pi, \pi_0, \sigma).
\end{equation}
It is an important advantage of Bayesianism that (3.2) makes sense as probability, whereas a frequentist meaning (if possible) would appear convoluted and unnatural. (3.2) is the appropriate Bayesian probability for use as a prior given unbiased measurements. Thus, if one first selects a star and then measures parallax then the probability for its true parallax depends only on the measurement, not on the distribution of other stars, and can be taken to be Gaussian with mean equal to the measured parallax. It is easy to show, by repeating the calculation of appendix A using $f(\pi|\pi_0)$ (3.2) in place of $g(\pi|\pi_0)$ (2.2), that $ E(\pi) = \pi' $. Thus, with a normal probability distribution for true parallax, it is found that the observed parallaxes are an unbiased estimator for the true parallax, whether calculated directly from the distribution of observations or from (3.2) interpreted as a Bayesian probability distribution.

\section{The Trumpler-Weaver bias}\label{sec:4}
The lower bound on the integrals in (2.5) has very little meaning, as we do not know the true parallax. This section will show that the consequence of choosing an inappropriate lower bound is that the L-K correction is a poor estimate of the Trumpler-Weaver bias, which affects a stellar samples selected according to a lower bound on observed parallax. A meaningful formula is 
\begin{equation}
\langle \Delta M \rangle = \frac{5  \int_{0}^{\infty}  \int_{\Pi}^{\infty} \pi^{-4} \log(\pi / \pi_0) N( \pi_0, \pi, \sigma) d\pi_0 d\pi}{\int_{0}^{\infty}  \int_{\Pi}^{\infty}  \pi^{-4} N( \pi_0, \pi, \sigma) d\pi_0 d\pi} ,
\end{equation}
where $ \Pi $ is the lower bound on observed parallaxes. This is essentially the formula West (1969) used to calculate the Trumpler-Weaver bias. LK1's argument attempts to short-cut the calculation of (4.1) by changing the lower bound on the integration, but I will show that this introduces an error. Substituting $ x = \pi/\sigma $, $ x_0 = \pi_0/\sigma $ we find 
\begin{equation}
\langle \Delta M \rangle = \frac{5  \int_{0}^{\infty}  \int_{\Pi/\sigma}^{\infty}  \pi^{-4} \log(x / x_0) N( x_0, x, \sigma) dx_0 dx}{\int_{0}^{\infty}  \int_{\Pi/\sigma}^{\infty}  \pi^{-4} N( x_0, x, \sigma) dx_0 dx} ,
\end{equation}
showing that the Trumpler-Weaver bias depends only on $ \sigma/\Pi $, the measurement error (assumed constant for a stellar class in a survey) as a fraction of the bounding parallax. 
\begin{figure}
	\centering
		\includegraphics[width=0.47\textwidth]{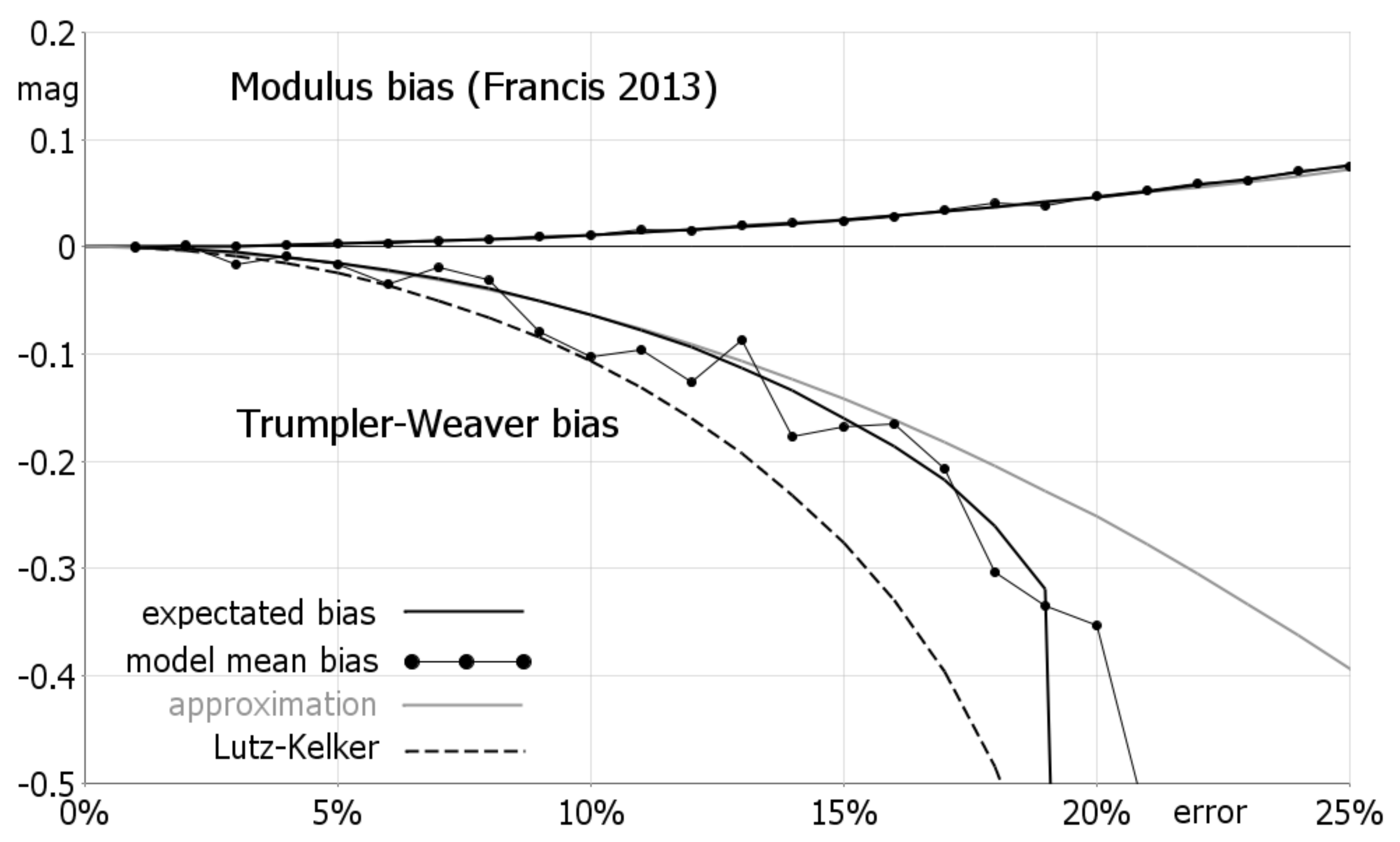}
	\caption{\textit{Negative bias:} The Trumpler-Weaver bias found from a: (continuous black) the expectation (4.2), b: (circles) using Monte Carlo simulation, c: (grey) quadratic approximation $ -6.3(\overline{\sigma/\pi_0})^2 $ with (dashed) numerical integration of the L-K formula (2.5). \textit{Positive bias:} Modulus bias (Francis 2013) from a: (continuous black) numerical integration, b: (circles) the same Monte Carlo model without a bound on parallax, c: (grey) the quadratic approximation  $ 1.15(\overline{\sigma/\pi_0})^2 $. }
	\label{Fig:1}
\end{figure}

Figure 1 shows the result of evaluating this formula numerically (continuous black line, downward curve). Also shown is the result of numerical integration of the L-K formula (2.5) (dashed) and the quadratic approximation (grey)
\begin{equation}
\Delta M = -6.5(\overline{\sigma/\pi_0})^2
\end{equation} 
This improves on the approximation, $ -5.8(\overline{\sigma/\pi_0})^2 $, given by Francis (2013), which relied only on a Monte Carlo simulation. The formula is independent of cut-off distance. The Trumpler-Weaver bias is estimated at $ -0.015 $ mag for parallax errors of $ 5 $\%, and $ -0.063 $ mag for parallax errors of 10\%. The L-K formula overestimates Trumpler-Weaver bias by more than 65\%. 

I checked the result using a Monte Carlo model with $ 30\,000 $ stars at each percentage point of the error in a uniform random distribution on a cube of side 1\,200 pc. A star with true distance $ R $ pc, has true parallax $\pi = 1\,000 / R $. I generated nominal measured parallaxes, $\pi_0$, by adding a random error to $\pi$ for each star, using a Gaussian distribution with standard error $ 0.01n\pi $ for integral values of $ n $. I then truncated the distribution using $ \pi > 10 $ and plotted the difference between the true and the observed sample magnitude for the model. The Monte Carlo model closely follows the predicted expectation calculated by numerical integration up to parallax errors below $ \sim 20\% $, at which point the prediction becomes pathological. The Monte Carlo model does not display pathological behaviour because it is necessarily based on a bounded distribution of stars.  

For comparison I also calculated the expected bias after replacing $g(\pi|\pi_0)$ (2.2) with $f(\pi|\pi_0)$ (3.2) in the L-K formula (2.5) (continuous black line, upward curve). This is a close match for the quadratic approximation (grey) given by Francis (2013) 
\begin{equation}
\Delta M = 1.15(\overline{\sigma/\pi_0})^2
\end{equation} 
I checked this formula by repeating the Monte Carlo method for the full population, i.e. without placing a lower bound on parallax. It is seen that the model almost exactly follows the expected prediction and explicitly confirms that $f(\pi|\pi_0)$ (3.2) is the correct Bayesian prior from which to calculate bias in magnitude resulting from errors in measurements of parallax for individual stars. These results show explicitly that bias due to parallax errors is positive and relatively small. The severe negative Trumpler-Weaver bias is due to population selection effects, not to the increase in stellar frequency at large radii, which is the same for both calculations.

\section{Conclusion}\label{sec:5}
It has been observed that the analysis of LK1 conflated the probability distribution for the true parallax of a star chosen randomly from a population to the probability distribution for the true parallax of a preselected star given a measurement of parallax. The use of a $ \pi^{-4} $ frequency distribution as a probability distribution contradicts the axioms of probability theory, according to which the integral of a probability density is normalised to unity. LK1 rectified this by introducing a lower bound on parallax, but this means only that the value of the lower bound is not important, not that there is no requirement for a lower bound. The statement that there is no lower bound is very different from the statement that a result is independent of the value of the lower bound. The use of the stellar frequency distribution in the manner of a Bayesian prior for the probability distribution of a single preselected star contradicts the assumption that measurement errors are Gaussian. This has implications for any study where the stellar distribution has been used as a Bayesian prior, not just LK1. For unbiased measurements with normal error distribution, the correct Bayesian probability distribution for true parallax given a measurement result is normal with mean equal to the measurement and standard deviation equal to the error. The bias due to errors in observed parallaxes is much smaller than, and of opposite sign to, the claimed L-K bias.

The question of the choice of an appropriate prior to compensate for selection effects, whether for an individual star or for a sample, must be considered separately from bias resulting from measurements. LK1 state in conclusion \textit{``Stars are not chosen for parallax measurement at random. If the parallax data are incomplete, a high probability exists that the stars which do have measured parallaxes were selected for some special reason. \ldots We have not yet attempted to investigate this point''}, but the calculation they gave assumes that stars are selected completely at random, without regard for either parallax or magnitude. This is unrealistic. It is inevitable for incomplete samples and individual stars that selection is influenced by magnitude, in which case a Malmquist correction should be made. If a comprehensive parallax survey is available then a cut on parallax together with a Trumpler-Weaver correction may be preferred because it is less affected by unknowns such as the magnitude distribution of a given stellar class.

Thus, contrary to LK1's claim that \textit{``the same type of systematic effect exists for all stars with observed trigonometric parallaxes''}, there is, in practice, no circumstance in which the L-K correction should be applied, either for individual stars, or for a predetermined sample of stars. The Trumpler-Weaver bias is a different type of bias, due to selection of a population according to parallax. A correction for the Trumpler-Weaver bias was calculated in section 4 and a useful quadratic approximation (4.3) was given. It was found that the L-K correction typically overcompensates the Trumpler-Weaver bias by more than 65\%. 

This has important implications for the interpretation of parallaxes from Hipparcos and Gaia, and for the calibration of the RR Lyrae zero point using parallaxes from HST. Benedict et al. (2011) applied a L-K correction, but according to the arguments here, this correction should not be applied. I recalculated zero points from the data, and found what appears to be a transcription error. Benedict applied an L-K correction to magnitudes given in Table 8, and plotted the L-K corrected fit for the $V$ band in Figure 8, but the figure for the zero point in the caption of Figure 8 and given in Table 9 is for the data without the L-K correction. Thus, for the $V$ band, there is close agreement with the reduced parallax method when the L-K correction is not used. The difference shown in Table 9 between zero points for the two methodologies for the $K$ band is in each case similar to the amount of the LK correction. Consequently, the figures from the reduced parallax method should be preferred.

\appendix
\section{The Expected parallax}\label{ap:A}
Assuming LK1's formula (2.2) for the probability distribution $g(\pi|\pi_0)$ of the true parallax given the observed parallax, a bias factor can be calculated for the expected parallax of a star with true parallax $\pi'$ by imposing a minimum value $\pi_{\mathrm{min}} $ on true stellar parallaxes.
\begin{equation}
\begin{split}
\frac{
E(\pi)}{\pi'} &= \frac{  \int_{\pi_{\mathrm{min}}}^{\infty}  \int_{-\infty}^{\infty} \pi g(\pi|\pi_0)f(\pi_0|\pi') d\pi_0 d\pi}{\pi'\int_{\pi_{\mathrm{min}}}^{\infty}  \int_{-\infty}^{\infty}  g(\pi|\pi_0)f(\pi_0|\pi') d\pi_0 d\pi}\\
&= \frac{ \int_{\pi_{\mathrm{min}}}^{\infty}  \int_{-\infty}^{\infty} \pi^{-3} N( \pi_0, \pi, \sigma) d\pi_0 d\pi}{ \pi' \int_{0}^{\infty}  \int_{-\infty}^{\infty} \pi^{-4} N( \pi_0, \pi, \sigma) d\pi_0 d\pi}\\
&= \frac{\sigma}{\pi'}\frac{ \int_{\pi_{\mathrm{min}}/\sigma}^{\infty}  \int_{-\infty}^{\infty} x^{-3} N( x_0, \pi' / \sigma, 1) dx_0 dx}{\int_{\pi_{\mathrm{min}/\sigma}}^{\infty} \int_{\Pi/\sigma}^{\infty} x^{-4} N( x_0, \pi' / \sigma, 1) dx_0 dx}\\
&\neq 1
\end{split}
\end{equation}
\begin{figure}
	\centering
		\includegraphics[width=0.47\textwidth]{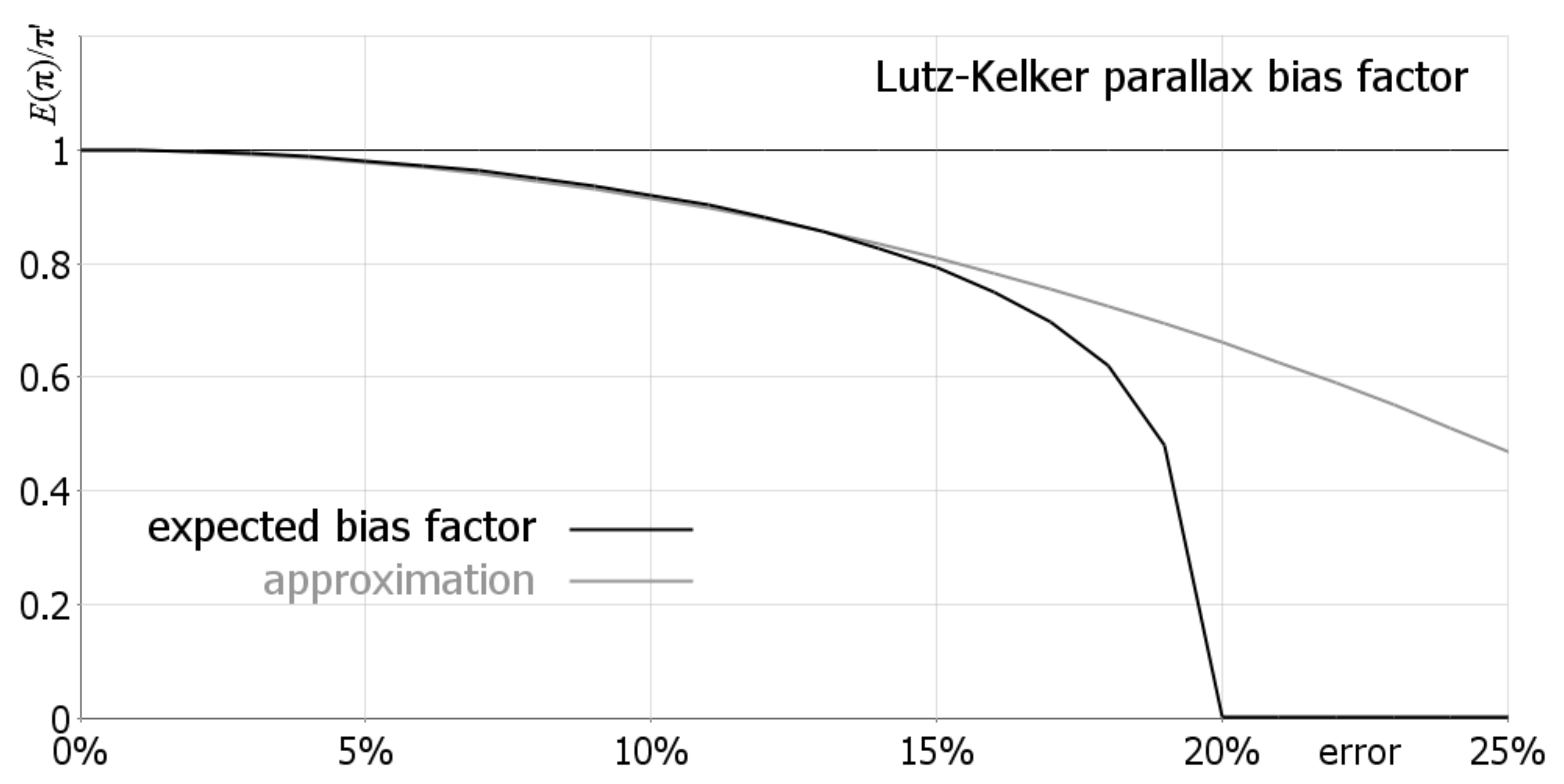}
	\caption{Bias factor calculated from the L-K probability distribution, with the approximation, $ 1 - 8.5(\sigma / \pi ')^2 $ shown in grey. }
	\label{Fig:2}
\end{figure}

The integrals can be evaluated numerically, shown in figure 2 together with the approximation, $ 1 - 8.5(\sigma/\pi')^2 $. The integrals in the numerator and denominator cannot be defined if a lower bound, $\pi_{\mathrm{min}} > 0 $, is not used, but the result converges and depends only on the proportional parallax error, not on $\pi'$ or $\pi_{\mathrm{min}}$. Thus the limit $\pi_{\mathrm{min}} \rightarrow 0 $ can be taken after calculating the bias factor. 

For a single pre-selected star, the result contradicts the assumption (2.1) that observed parallaxes are not biased. Thus this formula does not apply for the measurement of pre-selected stars. It applies to the mean parallax of a stellar sample selected according to a bound on true parallax, but this is not useful because the true parallax is not known and cannot be used as a selection criterion. For $ \sigma/\pi_{\mathrm{min}} > \sim20\% $ the bias factor is zero, showing that the level of bias is unmanageable for stellar samples containing stars at greater distances. This was also seen in figure 1. Replacing $g(\pi|\pi_0)$ (2.2) with $f(\pi|\pi_0)$ (3.2) in (A1) gives $E(\pi) = \pi' $ showing that observed parallax is an unbiased estimator if a normal probability distribution for true parallax is used.

\label{lastpage}

\end{document}